\documentclass[journal]{IEEEtran}
\usepackage{amsmath,amssymb,graphicx}
\usepackage{bm}             % bold, italic symbols
\usepackage{subcaption}
\renewcommand{\vec}[1]{\bm{#1}}
\DeclareMathOperator*{\argmin}{argmin}

\newcommand{\dens}{{\vec{\rho}}}
\newcommand{\dir}{{\vec{d}}}
\newcommand{\sca}{{\vec{s}}}
\newcommand{\tot}{{\vec{t}}}
\renewcommand{\ker}{{\vec{k}}}
\newcommand{\Ker}{{\vec{K}}}
\newcommand{\Abel}{{\vec{A}}}
\newcommand{\Prec}{{\vec{P}}}
\DeclareMathOperator{\diag}{diag}

\begin{document}
\title{Comparing One-step and Two-step\\ Scatter Correction and Density Reconstruction in X-ray CT}

\author{Alexander~N.~Sietsema, Michael~T.~McCann, Marc~L.~Klasky, and Saiprasad~Ravishankar%
\thanks{A. Sietsema is with the Dept. of Mathematics, Michigan State University, East Lansing, MI 48824, USA (sietsem6@msu.edu).}%
\thanks{S. Ravishankar is with the Dept. of Computational Mathematics, Science and Engineering and the Dept. of Biomedical Engineering, Michigan State University, East Lansing, MI 48824, USA (ravisha3@msu.edu).}%
\thanks{M. McCann and M. Klasky are with the Theoretical Division, Los Alamos National Laboratory, Los Alamos, NM 87545, USA (mccann@lanl.gov, mklasky@lanl.gov).}%
\thanks{This abstract has been accepted for poster presentation at the CT Meeting 2022.}%
}
\maketitle
\pagenumbering{gobble}
\begin{abstract}
In this work, we compare one-step and two-step approaches for X-ray computed tomography (CT)
scatter correction and density reconstruction.
X-ray CT is an important imaging technique in medical and industrial applications.
In many cases, the presence of scattered X-rays
leads to loss of contrast and undesirable artifacts in reconstructed images.
Many approaches to computationally removing scatter treat scatter correction as a preprocessing step that is followed by a reconstruction step.
Treating scatter correction and reconstruction jointly as a single, more complicated optimization problem is less studied.
It is not clear from the existing literature how these two approaches compare in terms of reconstruction accuracy.
In this paper, we compare idealized versions of these two approaches with synthetic experiments.
Our results show that the one-step approach can offer improved reconstructions over the two-step approach,
although the gap between them is highly object-dependent.
\end{abstract}


\IEEEpeerreviewmaketitle

\section{Introduction}
\label{sec:intro}
\IEEEPARstart{T}{he} presence of scattered X-rays presents a challenge for X-ray computed tomography (CT) imaging systems.
For example, in the context of medical cone-beam  CT,
scatter causes a loss in soft-tissue contrast and
artifacts such as
cupping, streaks, bars, and shadows~\cite{ruehrnschopf_general_2011}.
The same artifacts appear in nondestructive testing applications of X-ray CT,
where they can interfere with subsequent quantification tasks~\cite{lifton_experimental_2015}.
There is a large body of work on preventing scatter using hardware
and on correcting it using software;
see \cite{ruehrnschopf_general_2011,ruehrnschopf_general_2011a} for a review.

Hardware approaches to scatter correction include collimation
(blocking unwanted X-rays at the source,
 thereby preventing them from contributing to scatter)
 or increasing the distance between
 the source and detector,
 which reduces the amount of scattered radiation
 reaching the detector.
Among software approaches to scatter correction,
a key distinction is whether scatter correction happens
as a preprocessing step before CT reconstruction
or jointly with CT reconstruction.
In the former case, which we term \emph{two-step} reconstruction,
scatter is typically modeled as a function of the direct (i.e., not scattered)
radiograph.
This model is used to remove scatter from the measured data
and estimate the direct radiograph,
which is subsequently used for reconstruction.
In the latter case,
which we term \emph{one-step} reconstruction,
a model of scatter is included in a model-based CT reconstruction algorithm.

In our literature review, we found that
the two-step formulation is the more common approach,
including among recent work, e.g., 
\cite{bhatia_convolution_2017, maier_deep_2018,nomura_projection_2019,mccann_local_2021}.
A two-step approach is also implicitly assumed in works where only scatter correction is considered, e.g.,~\cite{tisseur_evaluation_2018}.
The popularity of the two-step approach may be because it usually involves solving two simple optimization problems
(scatter correction and reconstruction)
rather than a challenging joint problem.

The one-step approach appears less well-studied.
The model-based reconstruction formulation in~\cite{nuyts_modelling_2013} includes a scatter term,
but it is assumed to be known.
The authors in~\cite{mainegrahing_fast_2008} iteratively alternate between reconstruction and scatter estimation,
but do not formulate a joint optimization problem.
The review~\cite{ruehrnschopf_general_2011} describes a joint scatter correction and reconstruction approach,
but only in general terms and without implementing it.

The main goal of this work is to compare one-step and two-step reconstruction approaches
to find when, if ever, the added complexity of one-step reconstruction
yields better results.
To this end, we compare idealized one- and two-step algorithms on synthetic data in our studies.
These experiments are intentionally simple:
scatter is modelled using a convolution with a Gaussian kernel,
noise is Gaussian,
and the beam is monoenergetic;
we believe this experimental setup includes many of the key features of real descattering and reconstruction while leaving out aspects that may obscure the difference between the one-step and two-step approaches, e.g., inaccuracies in the radiographic forward model and beam hardening.

%%In a comparative experiment with simulated data, we show that one-step methods may outperform two-step methods (...)
In the following sections,
we formulate the scatter correction and reconstruction problem,
describe our one- and two-step algorithms,
and present our experiments, results, and conclusions. % TODO: outline correct?

\section{SCATTER CORRECTION AND RECONSTRUCTION}
We focus on a model of 2D monoenergetic X-ray tomography that includes scatter.
Given a (vectorized) density profile $\dens \in \mathbb{R}^{N_1 N_2}$,  our model of the total transmission $\tot$
is
\begin{equation}\label{eq:forwardmodel}
    \tot =  \sca + \dir, \quad 
    \sca = \Ker \dir, \quad
    \dir = \exp(-\xi \Abel \dens),
\end{equation}
where 
$\sca$ is the scattered signal,
$\dir$ is the direct (i.e., scatter-free) signal,
$\Abel\in \mathbb{R}^{M \times N_1 N_2}$ is the X-ray transform, $\xi$ represents the mass attenuation coefficient for a given material, 
$\exp(\cdot)$ is applied element-wise,
and $\Ker$ represents linear convolution by a kernel $\ker$, used to approximate scatter.

\begin{figure*}[htbp]
    \centering
    \begin{subfigure}{.3\linewidth}
    \centering
    \includegraphics[scale=0.3]{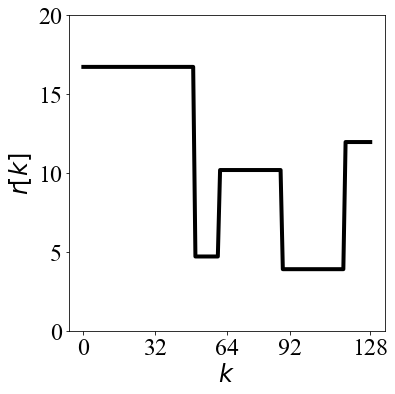}
    \caption{Density}
    \label{fig:denslineout}
    \end{subfigure}%
    \begin{subfigure}{.3\linewidth}
    \centering
    \includegraphics[scale=0.3]{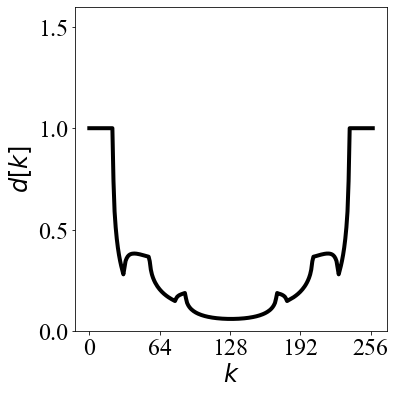}
    \caption{Direct}
    \label{fig:dirlineout}
    \end{subfigure}%
    \begin{subfigure}{.3\linewidth}
    \centering
    \includegraphics[scale=0.3]{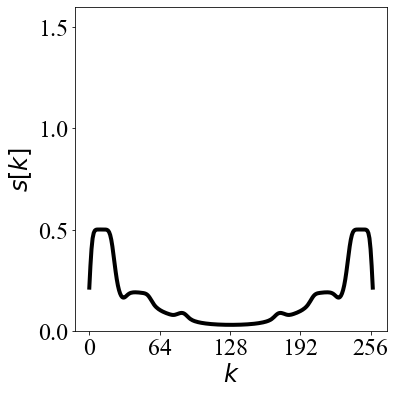}
    \caption{Scatter}
    \label{fig:scalineout}
    \end{subfigure}
    \caption{Density, direct, and scatter lineouts for profile 4 in our data set. }
    \label{fig:lineouts}
    
\end{figure*}

The choice to model scatter as a kernel convolved with the direct is common in the scatter correction literature~\cite{sun_improved_2010,mccann_local_2021, suri2006comparison, ohnesorge1999efficient, love1987scatter}.
This provides a fast scatter model that is at least representative of models used in practice. As our main goal is to bring out the differences between the one-step and two-step formulations, we leave more complicated models for future investigation.

\subsection{One-step vs two-step scatter correction and reconstruction}
\label{ssec:onesteptwostep}
\textbf{Two-step Approach:} The two-step method involves first
solving a scatter correction problem that is followed by solving a density reconstruction problem.
We formulate the first step, i.e. scatter correction, as
\begin{equation}
    \dir^* = \argmin_{\dir} \| \tot - (\Ker \dir  + \dir) \|_2^2 \label{eq:firststep},
\end{equation}
where we minimize an $\ell_2$ fit between the measured transmission and the model for it (i.e., direct + scatter) to account for noisy data. This step would correct for scatter in all the measured (one or multiple) CT views.

Following scatter correction,
we invert the nonlinear part of~\eqref{eq:forwardmodel} with the elementwise operation, $\dens_A^*[m, n] = -\log(\dir^*[m, n])/\xi$.
Note that we would need to set any values where $\dir^*[m, n]$ was less than or equal to zero to zero, as the logarithm would be invalid there.
We use the symbol $\dens_A^*$ here because this quantity represents the areal density~\cite{mccann_local_2021}.

After performing this scatter correction, we solve the following optimization problem to reconstruct the underlying density:
\begin{equation}
    \dens^* = \argmin_\dens \| \dens_A^* - \Abel \dens \|_2^2 + \alpha R(\dens), \label{eq:secondstep}
\end{equation}
where $R$ is a regularization functional and $\alpha$ is a nonnegative parameter. In essence, this approach first estimates the direct $\dir^*$ by removing scatter from the transmission $\tot$, and then takes that estimate and uses it to reconstruct an estimate of the object density $\dens^*$.

The main advantage of the two-step method is its simplicity:
both the first and second step are well-studied formulations of linear inverse problems that can be readily solved with standard algorithms.
The scatter correction step involves linear least squares (with possible constraints) for which there are many good algorithms, e.g., a fixed-point method such as Jacobi iteration~\cite{shewchuk_introduction_1994},  the conjugate gradient method~\cite{shewchuk_introduction_1994} or ADMM~\cite{boyd_distributed_2011}.
The reconstruction step can be solved efficiently using ADMM.

\textbf{One-step Approach:} One-step scatter correction and reconstruction involves
jointly optimizing over the entire forward model rather than first optimizing for the direct radiograph. This amounts to solving the following optimization problem:
\begin{equation}
    \dens^* = \argmin_\dens \| \tot - (\Ker + \vec{I}) \exp(- \xi \Abel \dens) \|_2^2 + \alpha R(\dens),\label{eq:onestep}
\end{equation}
where $R$ is again a regularization functional. Intuitively, the one-step method has the benefit of not relying on the estimation of the direct: in the two-step method, our overall estimate of the density is limited by the estimate of $\dir^*$ we obtain from solving~\eqref{eq:firststep}.
However, the one-step optimization could be more challenging depending on the complexity of the entire forward model.
%%% Mike, you should mention your example that these are actually solving different problems

\subsection{Implementation of one-step and two-step methods}
In our implementation of the one-step and two-step methods described above,
we consider imaging a spherically symmetric, single-material object,
parameterized by its radial profile $\dens \in \mathbb{R}^N$.
This simple model captures the important elements of X-ray reconstruction with scatter,
while remaining fast to optimize;
similar models find application in nondestructive testing~\cite{asaki_abel_2006}
and have been used for developing scatter correction methods~\cite{mccann_local_2021}.
As a result, our operator $\Abel$ is the forward Abel transform and is followed by spinning the 1D signal into a 2D image on which the convolution (for scatter) is applied.
For an example of data generated with this model, see Figure~\ref{fig:lineouts}.

For regularization, ($R$ in \eqref{eq:secondstep} and \eqref{eq:onestep}) we use total variation on the profile, i.e., $R(\dens) = \sum_{k=1}^n \left|\dens[k] - \dens[k-1]\right|$.

In both approaches, we solve the underlying optimization problems using the LBFGS algorithm in PyTorch~\cite{NEURIPS2019_bdbca288}. Note that PyTorch automatically handles non-differentiability issues.  Additionally, due to the poor conditioning of the Abel matrix $\Abel$, we use the separable quadratic surrogate (SQS) preconditioning as follows:
\begin{equation}\label{eq:preconditioner}
    \Prec = \diag\left( \Abel^\top \Abel \mathbf{1}\right)^{-1},
\end{equation}
where $\mathbf{1} \in \mathbb{R}^N$ denotes a vector of ones. We normalize $\Abel^\top \Abel \mathbf{1}$ prior to computing \eqref{eq:preconditioner}. This preconditioner is used for performing the one-step optimization and the second step of the two-step optimization. We apply this preconditioning 
%in the one-step and second step of the two-step algorithms; this is accomplished 
by premultiplying $\dens$ by $\Prec$ prior to optimizing. For the one-step method, this then becomes the preconditioned optimization formulation
\begin{multline}
    \dens' = \argmin_\dens \| \tot - (\Ker + \vec{I}) \exp(- \xi \Abel \Prec \dens) \|_2^2 \\+ \alpha \sum_{k=1}^n \left|\Prec \dens[k] - \Prec \dens[k-1]\right|\label{eq:onesteppreconditioned},
\end{multline}
where the solution is recovered via $\dens^* = \Prec \dens'$;
a similar formulation is used for the two-step problem.

\section{EXPERIMENTS AND RESULTS}
\label{sec:experiments}

We compare the one-step and two-step algorithms by scatter correcting and reconstructing ten synthetically generated transmissions.
In all experiments, we quantify performance by taking the root mean square error (RMSE) between the ground truth density and the reconstructed density for each algorithm.

\subsection{Data generation}
\label{ssec:datageneration}
To generate test objects, we first randomly selected indices in 1D at which shells start and end, assuming a maximum radius of $N=129$ pixels of the object.
We converted these indices into a piecewise-constant profile with steps at each shell boundary.
Finally, we picked shell densities (in $\mathbb{R}$) uniformly in the range (0, 20), and assigned each shell a density.
These 1D radial profiles were spun to create 2D images (representing slice of 3D volume) as part of the forward projection process \cite{gibson_pyabel/pyabel_2021}.
%Optimizing for the density in 1D instead of 2D helps in part to avoid center effects caused by the Abel matrix.
The spinning process slightly cropped the images to avoid edge effects. 

In order to generate transmissions, we applied the forward model in~\eqref{eq:forwardmodel}, choosing $\xi=1 \times 10^{-3}$ and fixing a Gaussian scatter kernel. In our implementation, this kernel is a three-fold convolution with a $7\times7$ Gaussian blur with standard deviation $1.5$ pixels. Spinning occurs after $\Abel \dens$ is computed. Finally, Gaussian noise is added in the end with $\mu=0, \sigma=3\times 10^{-2}$, and negative values in the transmission are set to zero (non-physical). See Figure~\ref{fig:lineouts} for an example of the density, direct, and scatter profiles. 
%for profile 4.

\subsection{Implementation details}
\label{ssec:implementation}

\begin{figure}%[htbp]
   \centering
   \begin{subfigure}{0.5\linewidth}
   \includegraphics[width=\linewidth]{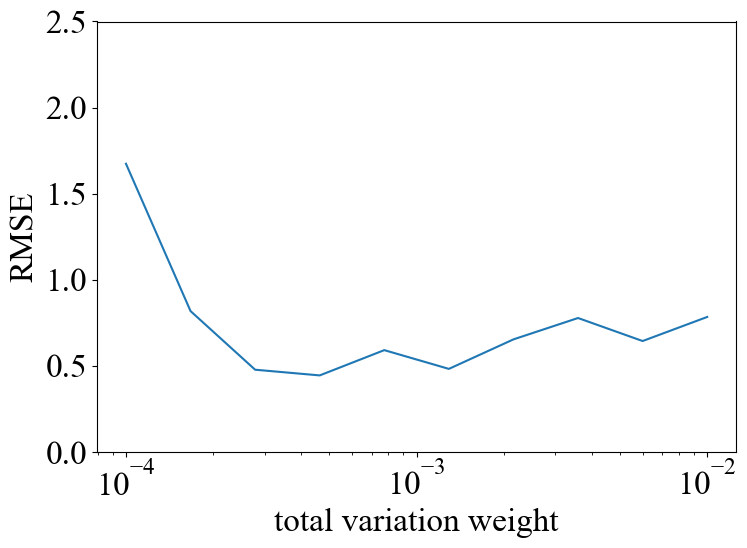}%
   \caption{One-step}
   \label{fig:onestepsweep}
   \end{subfigure}%
   \begin{subfigure}{0.5\linewidth}
   \includegraphics[width=\linewidth]{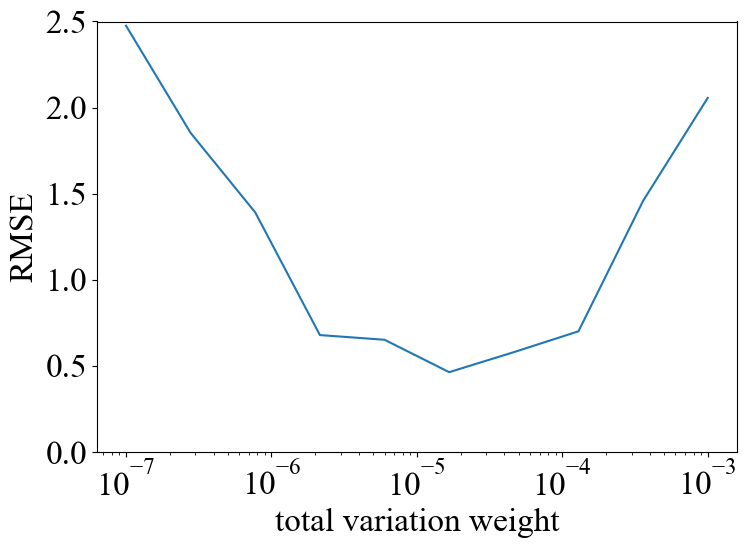}%
   \caption{Two-step}
   \label{fig:twostepsweep}
   \end{subfigure}
   \caption{Reconstruction accuracy (RMSE) as a function of the total variation weight ($\alpha$ in \eqref{eq:secondstep} and \eqref{eq:onestep}). The optimal learning rate was tuned independently for each weight value. The minima achieved are used in later fitting.}
   \label{fig:parametersweeps}
\end{figure}

All experiments were performed in Python using simulated data, and optimization is performed using the PyTorch package.
The one-step %model was fit 
algorithm was run
with a learning rate (step size of Wolfe line search) of $3 \times 10^{-2}$ and a total variation weight of $1 \times 10^{-3}$ with 20 iterations.
The two-step algorithm used a learning rate of $1.0$ and no total variation with 10 iterations for the first step, and a learning rate of $1 \times 10^{-2}$ and a total variation weight of $7 \times 10^{-4}$ with 20 iterations for the second step. 
%These constants 
All parameters
were optimized by performing a grid search over possible combinations of learning rate and total variation parameters.
See Figure~\ref{fig:parametersweeps} for the results of the grid search.
The forward Abel transform was the Hansenlaw method~\cite{hansen_recursive_1985} from the PyAbel Python package~\cite{gibson_pyabel/pyabel_2021}. 

% \begin{figure}%[htbp]
%   \centering
%   \begin{subfigure}{\linewidth}
%   \centering
%   \includegraphics[scale=0.3]{figures/onestep_parameter_sweep.png}
%   \caption{One-step}
%   \label{fig:onestepsweep}
%   \end{subfigure}\\
%   \begin{subfigure}{\linewidth}
%   \centering
%   \includegraphics[scale=0.3]{figures/twostep_parameter_sweep.png}
%   \caption{Two-step}
%   \label{fig:twostepsweep}
%   \end{subfigure}
%   \caption{Reconstruction accuracy (RMSE) as a function of the total variation weight ($\alpha$ in \eqref{eq:secondstep} and \eqref{eq:onestep}). The optimal learning rate was tuned independently for each weight value. The minima achieved are used in later fitting.}
%   \label{fig:parametersweeps}
%   \vspace{-0.2in}
% \end{figure}

\subsection{Results}
\label{ssec:results}
\begin{table}%[htbp]
\centering
\caption{One-step vs. two-step reconstruction performance.}
\begin{tabular}{|c|c|c|}\hline
     Profile & One-step RMSE & Two-step RMSE \\\hline\hline
        1 & \textbf{0.827} & 1.445\\
        2 & \textbf{2.077} & 9.242\\
        3 & \textbf{2.021} & 9.552\\
        4 & \textbf{0.568} & 2.468\\
        5 & \textbf{1.210} & 2.683\\
        6 & \textbf{2.488} & 3.071\\
        7 & 0.853 & \textbf{0.552}\\
        8 & \textbf{0.422} & 0.623\\
        9 & \textbf{1.887} & 2.397\\
        10 & \textbf{2.305} & 3.491\\
\hline
\end{tabular}

\label{table}
\end{table}

\begin{figure}[htbp]
    \begin{subfigure}{0.5\linewidth}
    \centering
    \includegraphics[width=\linewidth]{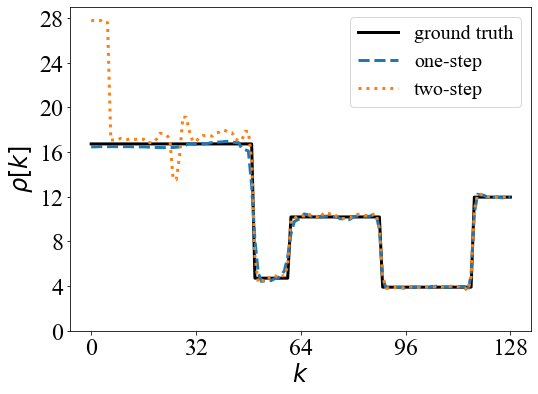}%
    \caption{Profile 4}
    \label{fig:prof4}
    \end{subfigure}%
    \begin{subfigure}{0.5\linewidth}
    \centering
    \includegraphics[width=\linewidth]{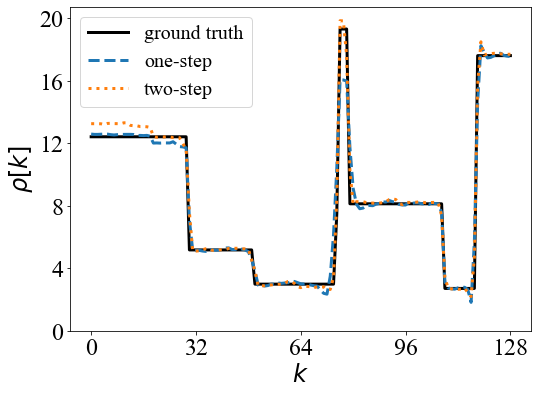}%
    \caption{Profile 7}
    \label{fig:prof7}
    \end{subfigure}
    \caption{Comparison of one-step and two-step reconstruction results on two profiles.}
    \label{fig:osvts}
\end{figure}

Results for each of the ten profiles are summarized in Table~\ref{table}.
Overall, the one-step algorithm outperforms the two-step algorithm, with a median RMS error of 1.548, compared to 2.575 for the latter. Two example reconstructions are shown 
in Figure~\ref{fig:osvts}. Qualitatively, the two-step method can be quite noisy, especially near the center of the reconstruction, despite the regularization. However, the two-step model may reconstruct thin shells better (Figure~\ref{fig:prof7}), although this benefit is on the whole negligible.

Both methods tend to struggle with profiles with high-density (densities above 8) shells near the center. This is likely due to the conditioning effects of the Abel matrix, which the SQS preconditioning is only partially able to solve. Uniformly high-density profiles  can also cause some problems in both methods (e.g., profiles 2 and 3).
This may be an artifact of our noise model,
as denser objects produce smaller transmissions,
which results in a low signal-to-noise-ratio when noise with a constant variance is added.

\begin{figure}
    \begin{subfigure}{0.5\linewidth}
    \centering
    \includegraphics[width=\linewidth]{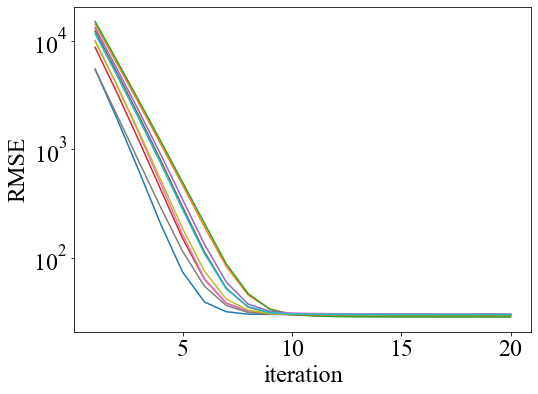}%
    \caption{One-step}
    \label{fig:osconvergence}
    \end{subfigure}%
    \begin{subfigure}{0.5\linewidth}
    \centering
    \includegraphics[width=\linewidth]{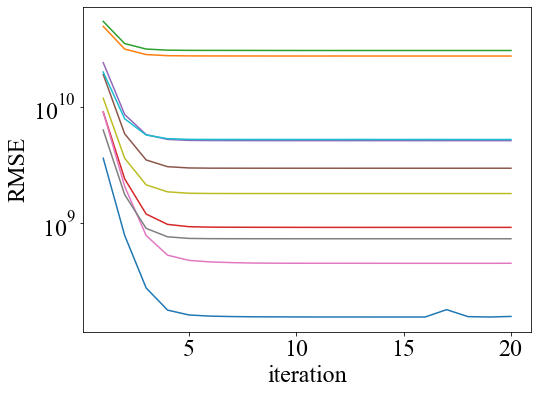}%
    \caption{Two-step (second step only)}
    \label{fig:tsconvergence}
    \end{subfigure}
    \caption{Comparison of one-step and two-step (second step only) convergence rates for each of the ten profiles.
    Satisfactory convergence is demonstrated in all cases. Note that the two-step loss is calculated in the areal density space, so its values are much higher than the one-step method's.}
    \label{fig:fullconvergence}
\end{figure}

Finally, we summarize convergence results, where we plot the data-fidelity terms in~\eqref{eq:secondstep} and ~\eqref{eq:onestep} over the LBFGS iterations.
From Figure~\ref{fig:fullconvergence}, we see that both algorithms converge quickly. We omitted the first step for the two-step method here since it converges nearly instantly and we plotted for only the one-step method and the second step of the two-step method. 

\section{CONCLUSIONS}
\label{sec:conclusion}
In this paper, we performed an empirical comparison of one-step and two-step X-ray CT scatter correction and reconstruction.
Our experiments showed that the one-step approach can demonstrate significant improvements over the %more 
common two-step approach.
While certainly not exhaustive, these experiments suggest that the added complexity of the one-step method may be worth it.
Our future work in this area would extend these experiments to more complicated regimes, including using more complicated scatter estimation models (e.g., \cite{mccann_local_2021}), polyenergetic spectra, and multiple materials.
The same comparisons could be also run on more realistic synthetic data
(e.g., generated from particle transport simulations as in \cite{mccann_local_2021})
to validate the efficacy of one-step descattering in a setting where the scatter model used during reconstruction does not perfectly match the scatter in the data.
Finally, we aim to run similar comparisons on real, experimental data.

%\pagebreak
\bibliographystyle{./IEEEbib}
\bibliography{refs_Mike,refs}

% Note that often IEEE papers with subfigures do not employ subfigure
% captions (using the optional argument to \subfloat[]), but instead will
% reference/describe all of them (a), (b), etc., within the main caption.
% Be aware that for subfig.sty to generate the (a), (b), etc., subfigure
% labels, the optional argument to \subfloat must be present. If a
% subcaption is not desired, just leave its contents blank,
% e.g., \subfloat[].

\end{document}